\input{aipcheck}

\documentclass[
    ,final            
  ]
  {aipproc}

\layoutstyle{6x9}

\begin{document}

\title{The role of AGN in the migration of early-type galaxies from the blue cloud to the red sequence}


\author{Kevin Schawinski}{
  address={Einstein Fellow}, 
  altaddress={Department of Physics, Yale University, New Haven, CT 06511, U.S.A.}, 
  altaddress={Yale Center for Astronomy and Astrophysics, Yale University, P.O. Box 208121, New Haven, CT 06520, U.S.A.}
}

\def\Chandra{\textit{Chandra}}
\def\XMM{\textit{XMM-Newton}}
\def\Swift{\textit{Swift}}

\def\OI{[\mbox{O\,{\sc i}}]~$\lambda 6300$}
\def\OIII{[\mbox{O\,{\sc iii}}]~$\lambda 5007$}
\def\SII{[\mbox{S\,{\sc ii}}]~$\lambda \lambda 6717,6731$}
\def\NII{[\mbox{N\,{\sc ii}}]~$\lambda 6584$}

\def\Ha{{H$\alpha$}}
\def\Hb{{H$\beta$}}

\def\NIIHa{[\mbox{N\,{\sc ii}}]/H$\alpha$}
\def\SIIHa{[\mbox{S\,{\sc ii}}]/H$\alpha$}
\def\OIHa{[\mbox{O\,{\sc i}}]/H$\alpha$}
\def\OIIIHb{[\mbox{O\,{\sc iii}}]/H$\beta$}

\def\Ebmv{E($B-V$)}
\def\LOIII{$L[\mbox{O\,{\sc iii}}]$}
\def\Ledd{${L/L_{\rm Edd}}$}
\def\LOIIIs4{$L[\mbox{O\,{\sc iii}}]$/$\sigma^4$}
\def\LOIIIMbh{$L[\mbox{O\,{\sc iii}}]$/$M_{\rm BH}$}
\def\Mbh{$M_{\rm BH}$}
\def\Msigma{$M_{\rm BH} - \sigma$}
\def\Ms{$M_{\rm *}$}
\def\Msun{$M_{\odot}$}
\def\Msunyr{$M_{\odot}yr^{-1}$}

\def\ergs{$~\rm ergs^{-1}$}
\def\kms{$~\rm kms^{-1}$}

\newcommand\aaps{{AAPS}}
\newcommand\aj{{AJ}}
\newcommand\araa{{ARA\&A}}
\newcommand\apj{{ApJ}}
\newcommand\apjl{{ApJ}}
\newcommand\apjs{{ApJS}}
\newcommand\aap{{A\&A}}
\newcommand\nat{{Nature}}
\newcommand\mnras{{MNRAS}}
\newcommand\pasp{{PASP}}

\begin{abstract}
We present a general picture of the ongoing formation and evolution of early-type galaxies via a specific evolutionary sequence starting in the blue cloud and ending in the low-mass end of the red sequence. This evolutionary sequence includes a Seyfert AGN phase in the green valley, but this phase occurs too late after the shutdown of star formation to be responsible for it. Thus, the bulk of black hole accretion in low-redshift early-type galaxies occurs in post-starburst objects, and not concurrent with star formation. On the other hand, a low-luminosity AGN phase switching on at an earlier stage when some star formation activity remains may be responsible for destroying the molecular gas reservoir fueling star formation. 
\end{abstract}

\maketitle


\section{Introduction}

When selected purely by morphology, the population of early-type galaxies in the local Universe includes a substantial subset of objects that are not passively evolving. These active early-type galaxies may host current or recent star formation as well as ongoing black hole growth. Due to their nature, these active early-type galaxies represent a window onto the ongoing formation and evolution of spheroidal galaxies and the connection between star formation and black hole growth (\cite{2006Natur.442..888S, 2007MNRAS.382.1415S, 2009ApJ...690.1672S}).

\section{Migration from the blue cloud to the red sequence}

\begin{figure}[h]
  \includegraphics[angle=90, width=\textwidth]{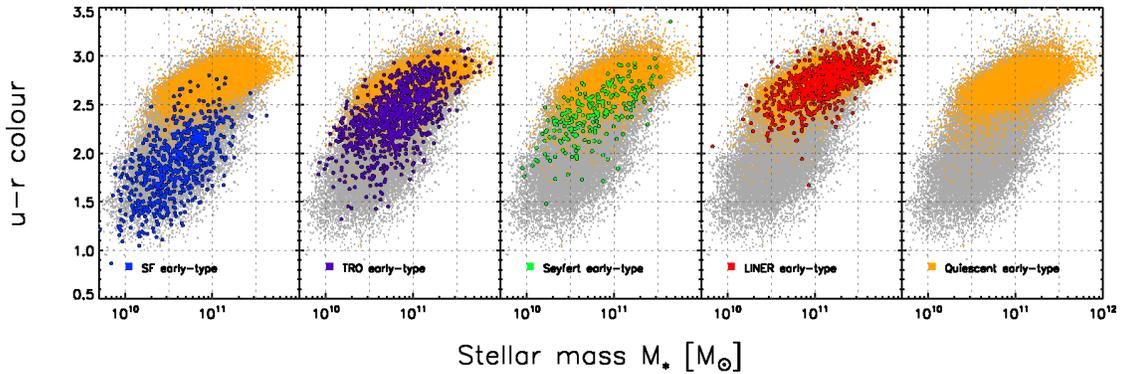}
  \caption{Color-mass diagrams for the MOSES sample of early-type galaxies (\citet{2007MNRAS.382.1415S}). In each panel, the gray points are the non-early-type galaxy population while the orange points are quiescent early-type galaxies without detectable emission lines. The colored points in each panel are early-type galaxies with emission lines, separated by diagnostic diagrams (\cite{1981PASP...93....5B, 2001ApJ...556..121K, 2007MNRAS.382.1415S}). The star-forming early-types (blue) reside in the blue cloud. the AGN+SF composites (TRO) have similar masses, but redder optical colors. The Seyfert AGN are redder yet. Some LINERs have similar masses and even redder colors, while the majority of them reside in massive red sequence early-type galaxies. \label{fig:ur_cmstellar}}
\end{figure}

SDSS reveals a strong association between the dominant source of gas ionization determined via emission line ratio diagrams and galaxy properties (e.g., star formation, Seyfert AGN, LINERs, AGN+SF composites; e.g., \citealt{1981PASP...93....5B, 2001ApJ...556..121K}). \citet{2007MNRAS.382.1415S} showed that the active early-type galaxy population whose nebular emission lines are dominated by star formation only cluster strongly in the blue cloud at relatively low stellar masses (see Figure \ref{fig:ur_cmstellar}). Early-type galaxies where AGN and star formation have roughly equal impact on the ionized gas occupy roughly the same mass range, but have redder optical colors. Redder yet, at the same mass, are the Seyfert AGN -- they perfectly occupy the green valley. LINERs cluster predominantly in massive red sequence early-type galaxies, but at the low masses characteristic of the other active early-types, they straddle the low mass end of the red sequence. 

If we take $u-r$ optical color as a proxy for stellar age, then this observational picture implies an \textit{evolutionary time sequence} for low-mass early-type galaxies starting from star formation in the blue cloud followed by the commencement of black hole accretion accompanied by a decline in star formation rate resulting in an optically `green' AGN host galaxy. As the underlying stellar population ages further, the galaxy then settles onto the low-mass end of the red sequence, exhibiting a LINER phase driven most likely by post-AGB stars (e.g., \cite{2008MNRAS.391L..29S, sarzi_gaspaper2}) for a while before settling into passive evolution on the red sequence.

\begin{figure}[h]
  \includegraphics[angle=90, height=.3\textheight]{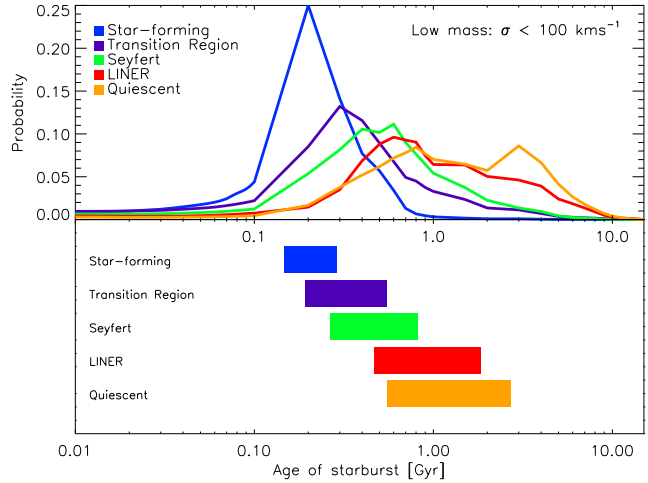}
  \caption{The evolutionary sequence for low-mass early-type galaxies with velocity dispersions less than 100\kms. Early-type galaxies pass through these phases on their way from the blue cloud to the red sequence. Note that the pure Seyfert AGN phase occurs several hundred Myr after the end of star formation (from \cite{2007MNRAS.382.1415S}).  \label{fig:sequence}}
\end{figure}

However, the interpretation of $u-r$ optical color as an age proxy is not unique. There are many plausible star formation histories that could give rise, for example, to the green optical colors seen in Seyfert AGN host galaxies that are in no way connected to either the star-forming blue early-types on the one hand, or the most recent arrivals on the low mass end of the red sequence on the other. More detailed work on the star formation histories of objects along this putative sequence is required to establish this evolutionary link, e.g. studies as performed by \citet{2007MNRAS.382.1415S}. By taking into account information from the UV-optical-NIR broad-band SED (from \textit{GALEX}, SDSS and 2MASS, respectively) and the stellar absorption (Lick) indices, they showed that the active early-type galaxies at fixed (low) mass (excluding the high-mass LINERs) all share the same recent star formation history: they all experienced a substantial recent burst of star formation in which 1--10\% of the stellar mass was formed; the only difference in objects with different emission line classifications is the time elapsed since this burst. Marginalizing over all nuisance parameters to recover the typical age of each phase yields the time sequence shown in Figure \ref{fig:sequence}.

\section{Do AGN suppress star formation?}

The evolutionary sequence presented in Figure \ref{fig:sequence} challenges scenarios where a luminous AGN phase is responsible for the shutdown of star formation by destroying the gas reservoir fueling the burst of star formation, as the Seyfert AGN phase occurs several hundred million years \textit{after} the end of substantial star formation (see also \cite{2009ApJ...692L..19S}). Seyfert AGN in early-type galaxies always occur in galaxies with post-starburst stellar populations. This observation rules out the Seyfert AGN as the agent that transforms blue early-type galaxies into red sequence objects. However, the AGN appears as a low-luminosity object at earlier times in the AGN+SF phase. Is this the phase along the evolutionary sequence where star formation is shut down?  

\citet{2009ApJ...690.1672S} observed the CO ($1\rightarrow 0$) and CO ($2 \rightarrow 1$) transition using the IRAM 30m telescope in a sample of objects along the evolutionary sequence to measure the amount of cold molecular gas present at each phase. They find that the star-forming early-types at the start of the sequence have ample molecular gas reservoirs of $\sim 10^{9}$\Msun, while all four Seyfert AGN do not yield any detections. It is within the AGN+SF phase that the molecular gas reservoirs are destroyed. In fact, the drop in molecular gas mass is so rapid that it cannot be accounted for by gas consumption due to star formation alone assuming the Schmidt law. The evidence thus points to an additional process active during the AGN+SF phase destroying the molecular gas reservoir and thus suppressing star formation. If we identify this process with the action of the AGN, then perhaps the low-luminosity AGN phase in the AGN+SF composites is radiatively inefficient, and therefore allows the remaining star formation to be visible in the emission lines, and does most of its work in a kinetic mode as a RIAF/ADAF (radiatively inefficient accretion disk/advection dominated accretion flow; \cite{1994ApJ...428L..13N}).

\section{Summary}

We have presented a general picture of the ongoing formation and evolution of early-type galaxies via a specific evolutionary sequence starting in the blue cloud and ending on the low-mass end of the red sequence. This evolutionary sequence includes a Seyfert AGN phase in the green valley, but this phase occurs too late after the shutdown of star formation to be responsible for it. Thus, the bulk of black hole accretion in low-redshift early-type galaxies occurs in post-starburst objects, and not concurrent with star formation. On the other hand, a low-luminosity AGN phase switching on at an earlier stage when some star formation activity remains may be responsible for destroying the molecular gas reservoir fueling star formation. 



\begin{theacknowledgments}
Support for the work of KS was provided by NASA through Einstein Postdoctoral Fellowship grant number PF9-00069 issued by the Chandra X-ray Observatory Center, which is operated by the Smithsonian Astrophysical Observatory for and on behalf of NASA under contract NAS8-03060. KS gratefully acknowledges support from Yale University.
\end{theacknowledgments}



\bibliographystyle{aipproc}   


\end{document}